\documentclass{elsart}
\def\ohalf{{\textstyle{1\over 2}}}

\newcommand{\beq}{\begin{equation}}
\newcommand{\eeq}{\end{equation}}
\newcommand{\beqa}{\begin{eqnarray}}
\newcommand{\eeqa}{\end{eqnarray}}

\usepackage{epsfig}

\begin{document}
\begin{frontmatter}
 \title{The role of $qqqq\bar{q}$ components in the nucleon and
the $N(1440)$ resonance}

\author[bar]{B. Juli\'a-D\'{\i}az} \,,
\author[hel]{D.O. Riska}

\address[bar]{Departament de Estructura i Constituents de la Materia,\\
Universitat de Barcelona, Diagonal 647, 08028 Barcelona, Spain}

\address[hel]{Helsinki Institute of Physics and Department of Physical
Sciences, P.O.Box 64, \\00014 University of Helsinki, Finland}

\date{}

\begin{abstract} 
The role of $q\bar q$ components in the nucleon and the $N(1440)$ 
resonance is studied by explicit coupling of the lowest positive 
parity $qqqq\bar q$ state to the $qqq$ components in the harmonic 
oscillator quark model. The lowest energy $qqqq\bar q$ component, 
where the 4-quark subsystem has the flavor-spin symmetry 
$[4]_{FS}[22]_F[22]_S$, is close in energy to the lowest positive 
parity excitation of the nucleon in the $qqq$ quark model. The 
confining interaction leads to a strong mixing of the $qqqq\bar q$ 
system and the positive parity excited state of the $qqq$ system.
This result is in line with the phenomenological indications for a 
two-component structure of the $N(1440)$ resonance. The presence 
of substantial $q\bar q$ components in the $N(1440)$ can bring about 
a reconciliation of the constituent quark model with the large 
empirical decay width of the $N(1440)$.
\end{abstract}
\end{frontmatter}

\section{Introduction}

The inadequacy of the conventional 3-quark model for the nucleons 
and the nucleon resonances in general is evident in its substantial 
underprediction of the values of both the electromagnetic and the 
strong decay widths of the lowest nucleon resonance, 
$N(1440)$~\cite{bruno,melde}. This indicates that this resonance 
should have significant $(q\bar q)^ n$, if not more exotic components, 
in addition to the 3 valence quarks. This is also suggested by the 
fact that this resonance appears naturally at its very low energy 
as a vibrational - i.e collective - state in the Skyrme model~\cite{bled}.

The admixture of $qqqq\bar q$ components in the nucleon and the 
$N(1440)$ resonance is studied here by explicit consideration of 
the coupling between the positive parity $qqqq\bar q$ configuration, 
which is expected to have the lowest energy, and the ground state 
and the lowest excited positive parity $qqq$ states. The main coupling 
between the $qqq$ and the $qqqq\bar q$ state is assumed to arise 
from the confining interaction. 

In the harmonic oscillator quark model, with color coupled confinement, 
the lowest energy positive parity $qqqq\bar q$ configuration will have 
an energy that falls fairly close to the lowest positive parity excited 
state of the $qqq$ configuration, if the hyperfine interaction depends 
both on spin and on flavor. In this case the lowest $qqqq\bar q$ configuration 
has the mixed flavor-spin symmetry $[4]_{FS}[22]_F[22]_S$. If the confining 
interaction couples as a Lorentz scalar, it leads to a strong coupling 
between the $qqq$ and the $qqqq\bar q$ configurations with positive parity. 
This coupling, which involves $q\bar q$ pair annihilation and creation, 
leads to significant mixing of the lowest energy configuration of the 
$qqqq\bar q$ system and the lowest positive parity excited state configuration 
of the $qqq$ system. With phenomenologically realistic parameter values this 
leads to a 2-state configuration, with strongly mixed $qqq$ and $qqqq\bar q$ 
components, for the $N(1440)$ resonance, which is in line with recent 
phenomenological analyses~\cite{Morsch}.

This article is arranged as follows: The $qqqq\bar q$ wave functions of 
the proton and the $N(1440)$ are described in section II. In section III the 
harmonic oscillator Hamiltonian for the $qqq$ and $qqqq\bar q$ configurations 
is defined along with the schematic hyperfine interaction model. The 
mixing of the states that is derived by the diagonalization of the 
Hamiltonian is described at the end of the section. In section IV the model 
is extended to the $qqq(q\bar q)^2$ system. Section V contains a 
concluding discussion. 

\section{The wave functions of the nucleon and the $N(1440)$}

In the harmonic oscillator quark model the wave functions of the nucleon 
and the $N(1440)$ in the $qqq$ configuration take the forms:
\beqa
\vert N,s_z \rangle_{3q}&=&{1\over\sqrt{2}}~\bigg[
\vert \ohalf,t_z\rangle_+
\vert \ohalf,s_z\rangle_+ +
\vert \ohalf,t_z\rangle_-
\vert \ohalf,s_z\rangle_-\bigg]
\varphi_{000}(\vec\xi_1)\varphi_{000}(\vec \xi_2),\nonumber\\
\vert N(1440),s_z \rangle_{3q}&=&{1\over\sqrt{2}}~\bigg[
\vert \ohalf,t_z\rangle_+
\vert \ohalf,s_z\rangle_+ +
\vert \ohalf,t_z\rangle_-
\vert \ohalf,s_z\rangle_-\bigg] \nonumber\\
&\times &{1\over\sqrt{2}}~\bigg(\varphi_{200}(\vec\xi_2)\varphi_{000}
(\vec \xi_1)
+\varphi_{000}(\vec\xi_2)\varphi_{200}(\vec \xi_1)\bigg).
\label{3qwave}
\eeqa
Here $\vert \ohalf,s_z\rangle_\pm$ and $\vert \ohalf,t_z\rangle_\pm$, 
are the spin and isospin wave functions, respectively, of mixed symmetry 
($[21]$), in which (+) denotes a state that is symmetric ``(112)'' and 
(-) denotes a state that is antisymmetric ``(121)'' under exchange of 
the spin or isospin of the first two quarks. The radial coordinates 
$\vec \xi_1$ and $\vec \xi_2$ are the Jacobi coordinates defined by 
the constituent coordinates as the first two of the general
Jacobi coordinates for a 
system of $n$ constituents:
\beqa
\vec\xi_k &=&{1\over \sqrt{k+k^2}} \left 
[\sum_{l=1}^k \vec r_l  - k \vec  r_{k+1} \right],  
\quad  k=1,...,n-1, \nonumber \\
\vec R &=&{1\over n}\sum_{l=1}^n \vec r_l  \, .
\label{Jacobi}
\eeqa

The two harmonic oscillator wave functions above are defined as:  
\beq
\varphi_{000}(\vec\xi)=\left({\tilde\omega_3^2\over \pi}\right)^{{3\over 4}}
~ e^{-\tilde\omega_3^2 \xi^2/2},
\qquad 
\varphi_{200}(\vec\xi)=\sqrt{{2\over 3}}\tilde\omega_3^2
\left(\xi^2-{3\over {2\tilde\omega_3^2}}\right)\varphi_{000}(\vec\xi)\, .
\label{harm3q}
\eeq
Here $\tilde\omega_3=\sqrt{m\omega_3}$, where $m$ is the constituent 
mass and $\omega_3$ is the oscillator frequency. The subscripts denote 
the quantum numbers $(nlm)$ of the oscillator wave functions.

Positive parity demands that the lowest energy $qqqq\bar q$ admixtures 
in the proton and the $N(1440)^+$ have to be $P-$wave states. By the 
conventional assumption that the hyperfine interaction between the
quarks is spin-dependent, it follows that the $qqqq$ configurations 
that have lowest energy are those for which the spin state has the 
highest possible degree of antisymmetry~\cite{helminen}. A similar 
argument applies in the case the hyperfine interaction is flavor 
dependent. The simplest hyperfine interaction model, which leads 
to a realistic splitting between the nucleon and the $N(1440)$, 
is the schematic flavor and spin dependent hyperfine interaction 
between the quarks 
$-C_\chi\sum_{i<j} \vec\lambda_F^i\cdot
\vec\lambda_F^j\,\vec\sigma_i\cdot\vec\sigma_j$, 
where $C_\chi$ is a constant parameter 
($C_\chi\sim 20-30$ MeV)~\cite{glozman}. 
This implies that the $qqqq\bar q$ configuration that has the lowest 
energy, and which is most likely to form notable admixtures in the 
nucleons and the $N(1440)$, has the mixed spin-flavor symmetry 
$[4]_{FS}[22]_F[22]_S$, with the antiquark in the ground 
state~\cite{helminen}.  

The wave function for this $qqqq\bar q$ component takes the form:

\begin{eqnarray}
\vert N,s_z \rangle_{5q} &=& 
{1\over\sqrt{2}}\sum_{a,b}\sum_{m,s} 
(1,\ohalf,m,s\vert\, \ohalf,s_z) 
C^{[1^{4}]}_{[31]_{a}[211]_{a}}\,
\nonumber\\
&&
[211]_C(a)\,[31]_{X,m}(a)\, [22]_F(b)\,[22]_{S}(b)\, 
\bar\chi_{t_z,s}\,\psi(\{{\vec \xi}_i\})\, .
\label{5qn}
\end{eqnarray}
Here the color, space and flavor-spin wave functions of the $qqqq$ 
subsystem have been denoted by their Young patterns respectively, 
and the sum over $a$ runs over the 3 configurations of the $[211]_C$ 
and $[31]_X$ representations of $S_4$, and the sum over $b$ runs 
over the 2 configurations of the $[22]$ representation of $S_4$ 
respectively~\cite{chen}. The symbols $C^{[1^{4}]}_{[31]_{a} [211]_{a}}$ 
denote the $S_4$ Clebsch-Gordan coefficients for the representations 
$[1111][31][211]$. The values of these coefficients are:
\beq
C_{[31]_1 [211]_1}^{[1^4]}={1\over \sqrt{3}}\, ,\qquad
C_{[31]_2 [211]_2}^{[1^4]}={-1\over \sqrt{3}}\, ,\qquad
C_{[31]_3 [211]_3}^{[1^4]}={1\over \sqrt{3}}\, .
\eeq
  
The explicit color-space part of the wave function~(\ref{5qn})
may be then expressed in the form:
\beqa
\psi_C(\{{\vec\xi}_i\})&=&{1\over \sqrt{3}}
\bigg\{[211]_3~\phi_{01m}(\vec\xi_1)\phi_{000}(\vec\xi_2)
\phi_{000}(\vec\xi_3)\nonumber\\
&-&[211]_2~\phi_{000}(\vec\xi_1)\phi_{01m}(\vec\xi_2)
\phi_{000}(\vec\xi_3) \nonumber \\ 
&-&[211]_1 ~\phi_{000}(\vec\xi_1)\phi_{000}(\vec\xi_2
)\phi_{01m}(\vec\xi_3)\bigg\}\phi_{000}(\vec\xi_4)\,.
\label{5qno}
\eeqa
Here the vectors $\vec\xi_i$, i=1,..,4, are the Jacobi coordinates 
for the 5 body system as in Eq.~(\ref{Jacobi}) 
and the oscillator wave functions for the quarks in the $qqqq\bar q$ 
system are:
\beqa
\phi_{000}({\vec\xi}_i)=\left({\tilde\omega_5^2\over \pi}\right)^{{3\over 4}}
~ e^{-\tilde\omega_5^2 \xi_i^2/2},~~
\phi_{01m}({\vec\xi}_i)=\sqrt{2}\tilde\omega_5 \xi_{i,m}\, 
\phi_{000}({\vec\xi_i}),
\label{harm5q0}
\eeqa
Here $\tilde\omega_5=\sqrt{m\omega_5}$, where $\omega_5$ is the 
oscillator frequency for the $qqqq\bar q$ system. The relation 
between the oscillator parameters for the $qqq$ and the $qqqq\bar q$ 
components depends on the color dependence of the confining interaction 
as shown below. In Eq.~(\ref{5qno}) $C_i$ ($i=1, 2, 3$) represent the 
color wave functions of the 3 configurations of $[211]_C$ and notice 
has been taken of the fact
that the vectors $\vec\xi_i$ ($i=1, 2, 3$) realize the 3 
configurations of $[31]_X$ in orbital space~\cite{chen}.

\section{Hamiltonian model}
\subsection{The non-interacting part}

The non-interacting part of the oscillator model Hamiltonians 
for the $qqq$ and $qqqq\bar q$ systems may be written in the 
following way:
\beqa
 H_3&=&3m + \sum_{i=1}^2\, \left\{ {\tilde \xi_i^2 \over 2 m} + 
{m\omega_3^2\over 2}\, \xi_i^ 2\right\}\, ,\label{h3q} \\
 H_5&=&5m + \sum_{i=1}^4\,\left \{ {\tilde \xi_i^2 \over 2 m} + 
{m\omega_5^2\over 2}\, \xi_i^ 2\right\}\, .\label{h5q}
\eeqa
Here $\tilde\xi_i$ denotes the momentum operator that is canonically 
conjugate to the position operator $\xi_i$ and $m$ denotes the 
constituent mass. These Hamiltonians are translationally and 
rotationally invariant, and therefore invariant under
Poincar\'e transformations as well, 
despite their non-relativistic appearance.

Hadron phenomenology suggests that the confining operator depends 
on color through the operator ${\bf\lambda}_i^C\cdot {\bf\lambda}_j^C$. 
If this color dependence is applied to the harmonic oscillator model 
for the confining interaction it implies a relation between the 
oscillator frequencies, $\omega_3$ and $\omega_5$, for the $qqq$ and 
$qqqq\bar q$ systems. 

The matrix elements of the color operator for any two quarks 
(or a quark and an antiquark) in the $qqq$ and $qqqq\bar q$ 
systems, which have the color symmetries $[1^3]$ and $[2^3]$ 
respectively, are~\cite{richard}:
\beq
<\lambda_i^C\cdot \lambda_j^C>_{[111]} = -8/3,\quad
<\lambda_i^C\cdot \lambda_j^C>_{[222]} = -4/3.
\eeq
From these matrix elements, one may infer the following relation 
between the oscillator frequencies for the $qqq$ and $qqqq\bar q$ 
systems~\cite{helminen}:
\beq
\omega_5 = \sqrt{{5\over 6}}\,\omega_3\, .
\eeq

The energies of the ground state and the first excited positive 
parity state of the $qqq$ system given by the Hamiltonian (\ref{h3q}) 
are
\beqa
&&E_3^0= 3m + 3\omega_3\, ,\\
&&E_3^2= 3m + 5\omega_3\, .
\eeqa
The lowest state of positive parity of the $qqqq\bar q$ 
Hamiltonian (\ref{h5q}) is
correspondingly \beq
E_5^0 = 5m + 7\omega_5\, . \label{e5q}
\eeq
In the absence of any hyperfine interaction it follows from these 
expressions 
that the energy of the 5-quark state falls $\sim 2 m + 1.5\omega_5$ 
(cf.(\ref{e5q})) above the excited positive parity state of the 
$qqq$ configuration. A sufficiently strong attractive
hyperfine interaction is of course required to lower the energy
of the lowest positive parity state of the $qqqq\bar q$ system
below that of the lowest negative parity state.

\subsection{The hyperfine interaction}

Consider now the situation in which the quarks (or the antiquarks) 
interact via 
the schematic flavor and spin dependent hyperfine interaction:
\beq
H_\chi=-C\sum_{i<j}\lambda_i^F\cdot\lambda_j^F
\vec\sigma_i\cdot\vec\sigma_j\, .
\label{hfi}
\eeq
The constant  $C\sim$ 30 MeV, represents an average of the matrix 
element of the hyperfine interaction in the low lying orbitals of 
the $qqq$ system. This interaction organizes the low lying part 
of spectrum of the $qqq$ system in a way that closely follows that 
of the empirical spectrum~\cite{glozman}. A dynamical
interpretation of this interaction may be given in terms
of single- and two-pion exchange between the quarks \cite{brown}.
As those interaction components lead to opposite signs for
the spin-spin interaction between quarks and antiquarks, with
a resulting at most weak spin-spin interaction, the hyperfine
interaction between quarks and antiquarks have been neglected here.
 
The matrix elements of this interaction in the $[3]_{FS}[21]_F[21]_S$
mixed flavor symmetry state of the conventional $qqq$ quark model
is $<H_\chi>_3=-14 C$~\cite{glozman}. The matrix element of the
hyperfine interaction~(\ref{hfi}) in the $qqqq$ configuration
with the mixed flavor symmetry $[4]_{FS}[22]_F[22]_S$ is 
$<H_\chi>_5 = -28 C$.

When this hyperfine interaction matrix element is taken into account, 
it modifies the energies of the $qqq$ and $qqqq\bar q$ states to: 
\beqa
&&E_3^{0'}= 3m + 3\omega_3 - 14 C\, ,\nonumber\\
&&E_3^{2'}= 3m + 5\omega_3 - 14 C\, ,\nonumber\\
&&E_5^{0'}= 5m + 7\omega_5 - 28 C\, .
\label{hfen}
\eeqa
This result shows that the energy of the lowest $qqqq\bar q$ state
with positive parity falls close to that of the lowest excited state
of the $qqq$ system, if $2m+1.5\omega_5 \simeq 14 C$. As $14 C\sim$ 
420 MeV, this condition may be fulfilled with phenomenologically 
realistic values for the constituent mass and the oscillator frequency.

Note finally, that if the coefficient $C$ in the interaction
Hamiltonian (\ref{hfi}), which represents the average strength
of the spatial dependence of the hyperfine interaction is allowed
to depend on the partial wave state of the quark pairs as in
\cite{glozman}, so that $C$ is replaced by state dependent
parameters $C_L$, the energy expressions (\ref{hfen}) are
modified to:
\beqa
&&E_3^{0'}= 3m + 3\omega_3 - 14 C_0\, ,\nonumber\\
&&E_3^{2'}= 3m + 5\omega_3 - 14 \left({C_0+C_2\over 2}\right)\, ,\nonumber\\
&&E_5^{0'}= 5m + 7\omega_5 - 28 \left({2C_0+C_1\over 3}\right)\, .
\label{hfenl}
\eeqa
These expressions will be employed below.

\subsection{Coupling between the $qqq$ and the $qqqq\bar q$ components}

\begin{figure}[t]
\begin{center}
\vspace{15pt}
\mbox{\epsfig{file=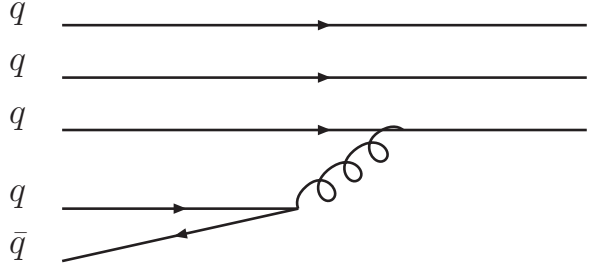, width=80mm}}
\caption{Confining mechanism responsible for the 
mixing between the $qqq$ and $qqqq\bar{q}$ states.
\label{fig1}}
\end{center}
\end{figure}

Consider then the coupling between the $qqq$ and the $qqqq\bar q$
states that is illustrated in Fig.~\ref{fig1}. If the mediating 
coupling is taken to be the confining interaction, which couples 
to the quarks and the antiquark as a Lorentz scalar, this coupling 
to lowest order in $p/m$ takes the form:
\beq
V_{cou}= 
-{1\over 2 m}
<\lambda_3^C\cdot\lambda_4^C>
\tilde v_{conf}(\vec p_4 + \vec p_5)\,
\vec\sigma_4\cdot(\vec p_5-\vec p_4)\, .
\label{pcoupling}
\eeq
Here $\tilde v_{conf}$ is the formal Fourier transform of the 
confining interaction. The spatial representation of the interaction 
is
\beq
V_{cou}= -{1\over 2 m}<\lambda_3^C\cdot\lambda_4^C>
 v_{conf}(\vec r_{34})\,\vec\sigma_4\cdot(\vec p_5-\vec p_4)\, .
\label{rcoupling}
\eeq 
Here $\vec r_{34}=\vec r_3 - \vec r_4$ and a $\delta$ function 
$\delta(\vec r_4-\vec r_5)$ is implied.  As above, in the 
$qqqq\bar q$ system $<\lambda_3^C\cdot\lambda_4^C> = -4/3$.

The explicit form for the confining coupling potential 
$ v_{conf}(\vec r_{34})$ is in the present harmonic oscillator 
model:
\beq
v_{conf}(\vec r_{34})= {1\over10}m\omega_5^2 r_{34}^2\, .
\eeq

In terms of the relative coordinates~(\ref{Jacobi}), one has:
\beqa
 r_{34}^2 &=& {2\over 3}(\vec\xi_2-\sqrt{2}\vec \xi_3)^2\, ,\\
 \vec p_5-\vec p_4 
&=& {1\over 2}(\sqrt{3}\tilde \xi_3-\sqrt{5}\tilde \xi_4) 
= -i {1\over 2} (\sqrt{3}\tilde \nabla _3 -\sqrt{5}\tilde \nabla _4)\, .
\eeqa
Here the $\tilde \nabla$ operators act on the corresponding relative 
coordinates $\vec\xi$. When these act on the incoming $P-$ state quark 
and $S-$state antiquark wave functions, the coupling interaction 
operator reduces to the form:
\beq
 V_{cou} = 
{2\sqrt{30}\over 375}\omega_5^2\sqrt{m\omega_5}(\vec \xi_2-\sqrt{2}\xi_3)^2
\sigma_{4,m} \left({\tilde\omega_5^2\over \pi}\right)^{3/2}
e^{-4\tilde\omega_5^2\xi_3^2/5}   \, .
\label{vcoun}
\eeq
The expression has been multiplied by an overall factor 4 to take into 
account the presence of 4 quarks in the $qqqq\bar q$ system. The matrix 
element of this interaction involves the matrix element over all 
coordinates of the $qqq$ system and the residual $qqq$ component of the 
$qqqq\bar q$ system. 

The spin matrix element gives rise to a factor $2\sqrt{3}$ and
the overlap of the color wave functions to a factor $1/\sqrt{3}$.
The final spatial matrix element for the coupling between the
ground states of the $qqq$ and the $qqqq\bar q$ systems then takes 
the form:
\beqa
<qqq\vert V_{cou}\vert qqqq\bar q> &=& 
{4\sqrt{30}\over 375}\omega_5^2
\sqrt{m\omega_5}
\left ({\tilde\omega_5^2\over \pi} \right)^{3/2}
\nonumber \\
&&\int d^3\xi_1d^3\xi_2d^3\xi_3\varphi_{000}(\vec\xi_1;\omega_3)
\varphi_{000}(\vec\xi_2;\omega_3)
\nonumber \\
&& (\vec\xi_2-\sqrt{2}\vec\xi_3)^2 e^{-4\tilde\omega_5^2\xi_3^2/5} 
\phi_{000}(\vec\xi_1;\omega_5)\phi_{000}(\vec\xi_2;\omega_5)\,.
\label{cou1}
\eeqa
This matrix element can be expressed in closed form as
\beq
V_c=<qqq\vert V_{cou}(\xi_3)\vert qqqq\bar q> =
{\sqrt{6}\over 5}m\left({\omega_3\over m}\right)^{3/2}
{\omega_5^3(5\omega_3+9\omega_5)\over
(\omega_3+\omega_5)^4}\, .
\label{matr1}
\eeq

The corresponding matrix element of the coupling interaction
between the positive parity excited state of the $qqq$ system 
and the ground state of the $qqqq\bar q$ system
is:
\beqa
&&<qqq^*\vert V_{cou}(\xi_3)\vert qqqq\bar q> =
{4\sqrt{15}\over 375}\omega_5^2
\sqrt{m\omega_5}
\left ({\tilde\omega_5^2\over \pi}\right)^{3/2}
\int d^3\xi_1d^3\xi_2d^3\xi_3 \nonumber\\ 
&&\{\varphi_{000}(\vec\xi_1;\omega_3)
\varphi_{200}(\vec\xi_2;\omega_3)
+\varphi_{200}(\vec\xi_1;\omega_3)
\varphi_{000}(\vec\xi_2;\omega_3)\}\nonumber\\
&& (\vec\xi_2-\sqrt{2}\vec\xi_3)^2 e^{-4\tilde\omega_5^2\xi_3^2/5} 
\phi_{000}(\vec\xi_1;\omega_5)\phi_{000}(\vec\xi_2;\omega_5)\, .
\label{vcou2}
\eeqa
This matrix element may also be expressed in closed form:
\beqa
&&V_c^*=<qqq^*\vert V_{cou}(\xi_3)\vert qqqq\bar q> =\nonumber\\
&&{\sqrt{2}\over 5}m\left({\omega_3\over m}\right)^{3/2}
{\omega_5^3(15\omega_3^2+20\omega_3\omega_5
-27\omega_5^2)\over
(\omega_3+\omega_5)^5}\, .
\label{matr2}
\eeqa
\subsection{Mixing between the $qqq$ and the $qqqq\bar q$ systems}

\begin{figure}[t]
\begin{center}
\vspace{25pt}
\epsfig{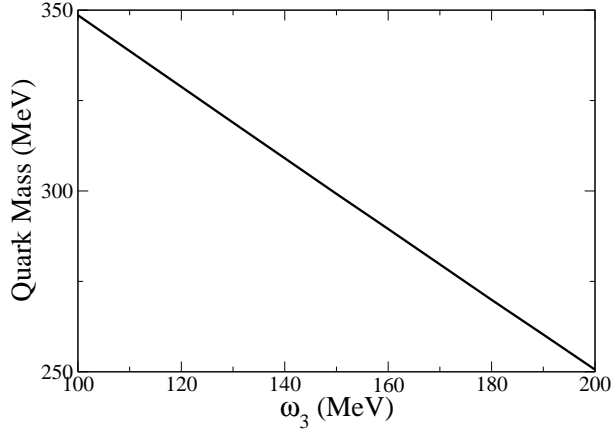}
\caption{The combinations of oscillator and constituent 
mass values, which yield 939 MeV for the lowest eigenvalue 
after diagonalization of the Hamiltonian (\ref{cchamil}).
\label{qmassvsw3}}
\end{center}
\end{figure}

The mixing between the $qqq$ and the $qqqq\bar q$ systems is determined by 
the $3\times 3$ Hamiltonian for the ground states of the $qqq$ and the 
$qqqq\bar q$ state and the excited positive parity state $qqq$ state. This 
Hamiltonian is
\beqa
H= \left(\matrix{  3m +3\omega_3 -14 C_0 & V_c     & 0     \cr
           V_c      & 5m + 7 \omega_5 -
28\bar{C}_{01}     & V_c^* \cr
         0                & V_c^*     & 3m+5 \omega_3 -14 \bar{C}_{02}}
 \right)\,. &&
\label{cchamil}
\eeqa
Here, $\bar{C}_{01}=(2C_0 + C_1)/3$ and $\bar{C}_{02}=(C_0 +C_2)/2$. 
Following ref.\cite{glozman} we set $C_0=29$ MeV, $C_1=45$ MeV
and $C_2=0$, for a good description of the low lying part 
of the baryon spectrum. By numerical diagonalization of this matrix, 
it is found that phenomenologically realistic eigenvalues for the 
3 states may be found by choosing the constituent quark mass $m$ 
to be in the range [250 $-$ 350] MeV and correspondingly the value 
for the oscillator parameter $\omega_3$ in the range [200 $-$ 100] MeV 
if the hyperfine interaction constants $C_L$ are chosen as above. 
The combinations of these parameters, which, upon diagonalization of 
the Hamiltonian (\ref{cchamil}), yield 939 MeV for the lowest eigenvalue 
are plotted in Fig.\ref{qmassvsw3}. The lowest eigenvalue obviously 
represents the nucleon.

\begin{figure}[t]
\begin{center}
\vspace{25pt}
\epsfig{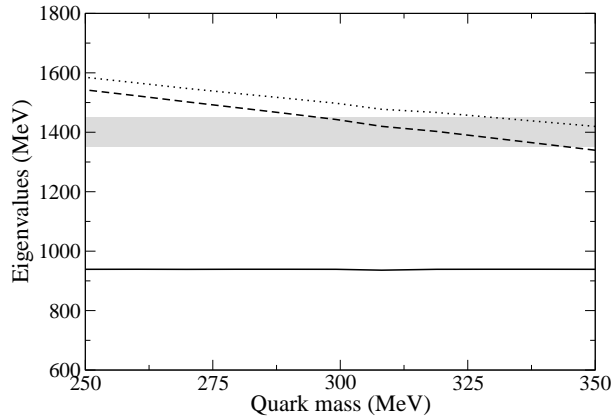}
\end{center}
\caption{\label{fig2}The eigenvalues of the Hamiltonian (\ref{cchamil})
as functions of the constituent mass. The band represents a 50 MeV interval 
around the PDG value for the Roper mass, 1440 MeV.}
\end{figure}

The corresponding 3 eigenvalues are shown in Fig.~\ref{fig2} as 
functions of the constituent mass. The two higher eigenvalues, 
which stay close together, may be interpreted as forming the broad 
$N(1440)$ resonance, for which there are clear phenomenological 
indications for a 2-state structure~\cite{Morsch}.  

\begin{figure}[t!]
\vspace{45pt}
\begin{center}
\epsfig{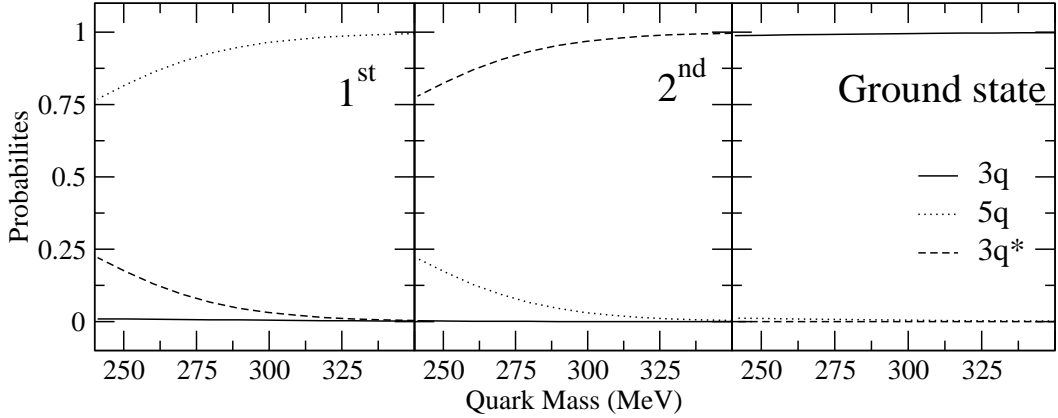}
\caption{\label{fig3}The probabilities of the 3 components in 
the 3 eigenstates as functions of the constituent mass. The
solid, dashed and dotted curves represent the $qqq$, $qqq*$ and
$qqqq\bar q$ components respectively }
\end{center}
\vspace{10mm}
\end{figure}

The probabilities of the $qqq$, the excited $qqq$ and the
$qqqq\bar q$ components in the 3 eigenstates are shown in
Fig.~\ref{fig3} respectively. The results indicates that 
in this model the nucleon is an almost pure $qqq$ state. In 
contrast there is a sizable mixing of the excited $qqq$ 
and the $qqqq\bar q$ components in the two close lying 
higher eigenstates if the quark mass lies below $\approx$275 MeV. 
The higher eigenstate is mostly made up of $qqqq\bar{q}$ with 
up to a 25$\%$ of $qqq^*$ state for a quark mass of $\approx$ 250 MeV. 
The middle eigenstate is, on the other hand, mostly made up of $qqq^*$ 
with up to $\approx 25\%$ component of $qqqq\bar{q}$ for a quark mass 
of $\approx$ 250 MeV. The admixture of the $qqqq\bar q$ component
in the ground state is mostly lower than $\sim$ 1 \% reaching 
this maximum value for a quark mass of 250 MeV.

\section{The $qqq(q\bar q)^2$ system}

While, as shown above, the ground state of the $qqqq\bar q$ 5-quark 
system may be expected to have an energy that falls within the vicinity 
of the lowest positive parity state of the $qqq$ system, the ground 
state of the $qqq(q\bar q)^2$ 7-quark system will have a much higher 
energy. This may be illustrated by a direct extension of the harmonic 
oscillator model employed above to the 7-quark system.

Consider again a harmonic confining interaction with the color 
dependence $\lambda_i^C\cdot\lambda_j^C$. The $qqq(q\bar q)^2$ system 
is a color singlet: $[333]$. The 5 quarks form a mixed symmetry color 
configuration: $[221]$, while the 2 antiquarks form the mixed symmetry 
color configuration $[211]$. The matrix elements of the color pair 
operator $\lambda_i^C\cdot\lambda_j^C$ in these configurations are:
\beq
<\lambda_i^C\cdot\lambda_j^C>_{[221]}=-{16\over 15}\, ,
\quad 
<\lambda_i^C\cdot\lambda_j^C>_{[211]}=-{4\over 3}\, .
\label{7color}
\eeq
Thus the strength of the confining interaction for a quark pair 
in the $qqq(q\bar q)^2$ system is $4/5$ times that of the $qqqq\bar q$ 
system, while the strength of the confining interaction between the 
2 antiquarks is the same as that between the quarks and the antiquarks 
in the $qqqq\bar q$ system.

From Eq.~(\ref{7color}) and the vanishing of the Casimir operator 
$C_2^{(3)}$ of SU(3) it follows that the matrix element of the 
operator $\lambda_i^C\cdot\lambda_j^C$ in a $q\bar q$ pair state 
in the $qqq(q\bar q)^2$ system is:
\beq
<\lambda_i^C\cdot\lambda_j^C>{q\bar q}=-{8\over 15}\, .
\label{qbarq}
\eeq 
Thus the strength of the confining interaction between a quark 
and an antiquark in the 7-quark system is one half of that between 
two quarks in the same system.

The oscillator Hamiltonian for the 7-quark system may now be 
written as:
\beqa
H&=&\sum_{i=1}^7 {p_i^2\over 2 m}-{P^2\over 14 m}
+{m\omega_{qq}^2\over 2}\sum_{i<j}^5 r_{ij}^2\, \nonumber\\
&&+{m\omega_{\bar q\bar q}^2\over 2} r_{67}^2
+{m\omega_{q\bar q}^2\over 2}\sum_{i=1}^5\sum_{j=6}^7 r_{ij}^2\,.
\label{7quarks}
\eeqa
Here $\vec P$ is the center-of-mass momentum.

This Hamiltonian is analytically diagonalized by first 
enlarging the system of relative coordinates for 5 quarks 
with the following two coordinates in Eq.~(\ref{Jacobi}).

The diagonalization is then achieved by introduction of 
the following new set of relative coordinates:
\beqa
\vec \eta_i&=&\vec\xi_i,\quad i=1,...4\, ,\\
\vec \eta_5&=& \sqrt{7\over 12}\vec \xi_5+\sqrt{5\over 12}\vec\xi_6\, ,\\
\vec \eta_6&=&-\sqrt{5\over 12}\vec \xi_5+\sqrt{7\over 12}\vec\xi_6\, .
\label{diag}
\eeqa 
The diagonalized version of (\ref{7quarks}) is then:
\beqa
&&H=\sum_{i=1}^6 {\tilde \eta _i^2\over 2 m}
+{m\over 2}(5\omega_{qq}^2+2\omega_{q\bar q}^2)\sum_{i=1}^4 \eta_i^2
\, \nonumber \\
&&+{7m\over 2} \omega_{q\bar q}^2 \,\eta_5^2+{m\over 2}
(2 \omega_{\bar q\bar q}^2 +
 5\omega_{q\bar q}^2)
\eta_6^2\, .
\label{Hdiag}
\eeqa
Here $\tilde \eta_i$ denotes the momentum operator that is
canonically conjugate to $\vec \eta_i$. In the absence of any 
coupling to the 3-quark or 5-quark system the energy of the ground 
state of this Hamiltonian is
\beq
E_7^0=7m + 6\sqrt{5\omega_{qq}^2+2\omega_{q\bar q}^2}
+{3\sqrt{7}\over 2}\omega_{q\bar q}+{3\over 2}\sqrt{2 \omega_{\bar q
\bar q}^2 + 5\omega_{q\bar q}^2}\, .
\label{7energy}
\eeq

Consider then the energy shift caused by the hyperfine interaction
(\ref{hfi}) between the quarks and between the 2 antiquarks in the
$qqqqq\bar q$ system. The lowest energy is that for the configuration 
which has the highest possible degree of antisymmetry in both the
flavor and the spin configurations. Moreover as positive parity
allows all the constituents in the 7-quark system to be in their 
ground state, the spatial state is totally symmetric. As a consequence 
the flavor-spin state is also totally symmetric as is the combined 
flavor-spin configuration.

In the ground state the flavor-spin configuration of the 
5 quarks in the 7 quark system is the mixed symmetry configuration
$[5]_{FS}[32]_F[32]_S$. The two antiquarks form the mixed
symmetry flavor-spin configuration $[2]_{FS}[11]_F [11]_S$. The matrix 
elements of the hyperfine interaction (\ref{hfi}) in these configurations
may be calculated from the general expression~\cite{helminen}:
\beq
<H_\chi>=-C\,\{ 4 C_2^{(6)} - 2 C_2^{(3)} - {4\over 3}
C_2^{(2)} - 8N\}\, ,
\label{casimir}
\eeq  
where $C_2^{(n)}$ is the quadratic Casimir operator of SU(n), and $N$ 
is the number of constituents.

This leads to the following energy shifts due to the hyperfine 
interaction in the 7-quark system:
\beq
<H_\chi>_{qqqqq}\, = - 40\, C\, ,\quad <H_\chi>_{\bar q\bar q}\,=-8\, C\, .
\label{7hfi}
\eeq

In order to estimate the ground state energy of the $qqq(q\bar q)^2$ 
system, the oscillator frequencies for the constituent pairs
in that system may be related through Eqs.~(\ref{7color}) 
and (\ref{qbarq}) to the oscillator frequency $\omega_3$ for the $qqq$ 
system. The resulting expressions are:
\beq
\omega_{qq}=\omega_3 \sqrt{ 2\over15}\, ,\quad
\omega_{q\bar q}={\omega_3\over \sqrt{15}}\, ,\quad
\omega_{\bar q\bar q}={\omega_3\over \sqrt{6}}\, .
\label{7freq}
\eeq
Insertion in (\ref{7energy}) with account of (\ref{7hfi}) yields:
\beq
E_7^{0'}=7 m +\omega_3 
\left(
{ 12\over \sqrt{5}} 
+{3\over 2}\sqrt{7\over 15}
+{3\over 2}\sqrt{{2\over 3}}\,\right)-48C\, .
\label{7inene}
\eeq
This leads to an approximate energy estimate, by which
the energy of the uncoupled 7-quark system falls 
$\sim 2m + 1.22\omega_3 -20 C \sim$ 200 MeV above that of the 5-quark 
system, and thus roughly 700 MeV above the $qqq$ ground state. Given 
the results in the previous section that only configurations that lie 
close in energy are strongly mixed,
the 7-quark system should then be expected to be strongly 
coupled only to higher lying baryon resonances than the $N(1440)$ 
(and the nucleon). Note that the form of the $q\bar q$ annihilation 
coupling couples the $qqq$ system to the $qqqq\bar q$ system 
and the $qqqq\bar q$ system to the $qqq(q\bar q)^2$ system, but
that there is no direct coupling of the $qqq$ and the $qqq(q\bar q)^2$ 
system, both of which have $L=0$ in their lowest positive parity 
configurations.

\section{Discussion}

This quark model study of the interplay between the lowest lying 
$qqqq\bar{q}$ component and the $qqq$ components of the nucleon 
and the N(1440) illustrates how the confining interaction mixes 
the $qqq$ and $qqqq\bar{q}$ components. The model exhibits the 
dual nature of the roper resonance, both as a genuine $qqq$ state 
and a $qqqq\bar{q}$ state. In the meson-baryon hadronic description 
it indicates the presence of a substantial $N\pi$ component in the 
$N(1440)$, for which there exist both theoretical arguments as well 
as experimental indications~\cite{Morsch,Krehl}. 

In the present model, the flavor-spin dependent hyperfine interaction 
singles out the most likely $qqqq\bar{q}$ configuration in the $N(1440)$. 
This 5q configuration lies close to the experimental value of the 
Roper resonances mass with phenomenologically reasonable values for the 
model parameters. Diagonalization of the model Hamiltonian for the ground 
state and the lowest positive parity excitation of the $qqq$ configuration
and the ground state of the $qqqq\bar q$ configuration leads to mixing 
between the lowest positive parity excited state of the $qqq$ configuration 
and the lowest energy $qqqq\bar q$ configuration. The configuration mixing 
ranges from 25\% for a quark mass of 250 MeV to 3\% for a quark mass of 300
MeV. Remarkably both states are close in mass for the considered 
range of quark masses. In the present model the $(q\bar q)$ admixture in 
the nucleon is very small, with a maximum value of $\sim$ 1\%.

The significance of the $qqqq\bar{q}$ component in the quark
model description of the electromagnetic transition form factors 
of the baryon resonances has recently been noted in Refs.~\cite{lix1}. 
This significance may also be indirectly inferred from the large
role played by meson-baryon in the coupled channel hadronic
model for the baryon resonances ~\cite{satolee}. With appreciable 
admixtures of $qqqq\bar q$ components in the resonances the 
failure of naive quark models to explain pionic decay widths
may also be overcome~\cite{lix2}. 

Finally it was found that admixtures of $(q\bar{q})^2 qqq$ 
components do not seem to be significant in the present model, 
if these states would have energies that are several hundred 
MeV above the lowest positive parity nucleon resonance.

\section*{Acknowledgment}
B.J-D. thanks the hospitality of the Helsinki Institute of 
Physics during the initiation of this work.

\end{document}